\documentclass[prb,showpacs,twocolumn]{revtex4}
\usepackage{graphicx,amssymb,amsmath}
\begin{document}
\title{Edge channel interference controlled by Landau level filling}
\author{L. V. Litvin }
\author{A. Helzel }
\author{H.-P. Tranitz }
\author{W. Wegscheider }
\author{C. Strunk}
\affiliation{ \mbox{Institut f\"{u}r experimentelle und angewandte
Physik, Universit\"{a}t  Regensburg, D-93040 Regensburg, Germany}}
\date{\today}
\begin{abstract}
We study the visibility of Aharonov-Bohm interference in an
electronic Mach-Zehnder interferometer (MZI) in the integer quantum
Hall regime. The visibility is controlled by the filling factor
$\nu$ and is observed only between $\nu \approx 2.0$ and $1.0$, with
an unexpected maximum near $\nu=1.5$. Three energy scales extracted
from the temperature and voltage dependences of the visibility
change in a very similar way with the filling  factor, indicating
that the different aspects of the interference depend sensitively on
the local structure of the compressible and incompressible strips
forming the quantum Hall edge channels.
\end{abstract}
%
%
\maketitle

\section{Introduction}
The electronic Mach-Zehnder interferometer\cite{ji} (MZI)  was
proposed to study the decoherence \cite{ji,marquardt} and orbital
entanglement effects \cite{sukh,neder2} by using edge channels in
the regime of the quantum Hall effect (QHE). Its high interference
contrast, observed at temperature of about 20 mK, is the consequence
of ballistic transport through the quasi-onedimensional edge
channels. The effect of temperature, \cite{ji,litvin,roull2} bias
voltage, \cite{ji,litvin,roull} and the interferometer size
\cite{neder,roull2} on the interference contrast are currently under
intense investigation. The magnetic field $B$ is another important
parameter, which controls the structure of the edge state consisting
of compressible and incompressible stripes. \cite{chklovskii} In
previous experiments both a monotonic growth of visibility with
increasing magnetic field \cite{neder} and a local maximum
\cite{roull2} were observed within the $\nu$=2 plateau. In these
works only the number of edge channels rather than the precise value
of the filling factor $\nu=nh/eB$ was specified. Here, $n$ is the
electron density, $h$ the Planck constant and $e$ the elementary
charge. In this paper, we systematically study the behavior of the
MZI visibility in a broad range of filling factors. We found that
the interference for the lowest Landau level (LL) appears at a
filling factor of $\nu$=2.0, reaches a maximum of visibility of
about 50~\% at $\nu$=1.5, and then decays to zero near $\nu$=1.0.
Although the interference occurs only in the outer edge channel, the
visibility is strongly affected by the presence of the inner edge
channel and its evolution, when $\nu$ is varied between 1 and 2.

\section{Experimental details}
%
\begin{figure}
\includegraphics[width=85mm]{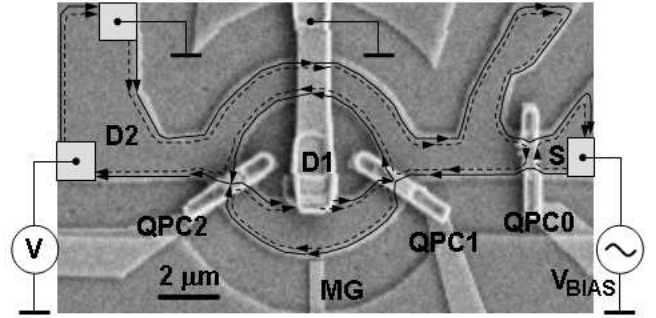}
\caption{SEM image of smaller Mach-Zehnder interferometer with the
scheme of edge states for filling factor 2. The transmissions of
QPC1 and QPC2 are set to 0.5. QPC0 reflects the inner edge channel
(dashed line) and transmits the outer one (solid line). Modulation
gate (MG) shifts the phase.}
\end{figure}
The interferometer (see Fig.~1) was fabricated on the basis of a
modulation doped GaAs/Ga$_{x}$Al$_{1-x}$As heterostructure
containing a two-dimensional electron gas (2DEG) 90~nm below the
surface. At~4~K, the unpatterned 2DEG density and mobility were
$n$=2.0$\times$10$^{15}$~m$^{-2}$ and $\mu$=206~m$^{2}/$(Vs),
respectively. Photolithography was employed to define Hall bars with
large contacts (connected to leads S, D2 and all gates at Fig.~1).
The ring-shaped interferometer mesa, contact D1, the quantum point
contacts  (with air bridges), and the modulation gate (labeled as MG
in Fig.~1) were patterned by means of electron-beam lithography. The
quantum point contact (QPC) No.0 was used to select the outer edge
channel for the interference experiment. We studied two MZIs with an
arm length of 14 and 9~$\mu$m and an arm width of $2.5-3 \mu$m ($1.7
\mu$m), respectively (see Fig.1). The QPCs of the larger MZI had
120~nm gap between sharp tips (tip radius $\approx$50~nm); for the
smaller MZI a 400~nm gap was used. The area between two interfering
paths, determined from the period of the AB oscillations in a
magnetic field, was 48 and 25 $\mu m^{2}$ for these MZIs. A standard
lock-in technique ($f\sim\;$300~Hz) with 1~$\mu$V excitation at
terminal S and detection at terminal D2 was employed (see Fig.1).
Most of the measurements were performed at a temperature of 25~mK.

%
\begin{figure}
\includegraphics[width=85mm]{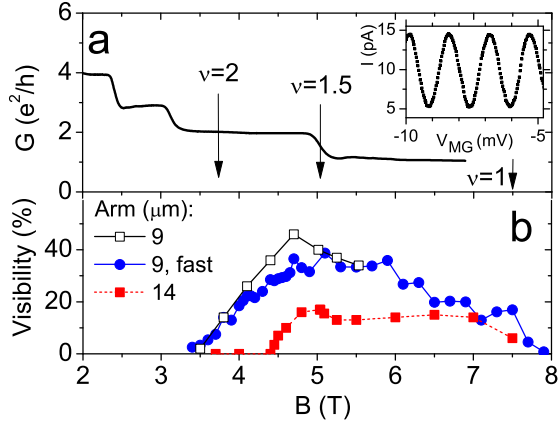}
\caption{(a) Two terminal magnetoconductance of the large
interferometer with all QPCs opened. Inset shows an example of
interference pattern at $B$=4.7T. (b) Visibility versus magnetic
field for both interferometers.  Full and open squares correspond to
data acquired with at least 4~h waiting time, full circles with 1~h.
}
\end{figure}

\section{Magnetic field dependence of the visibility}
In Fig.2 (inset) we show a typical trace of current $I(V_{MG})$ in
detector D2 vs~the voltage at the modulation gate $V_{MG}$. The
measured interference contrast is quantified by the visibility which
is defined as
\mbox{$\nu_{I}~=~(I_{max}-I_{min})/(I_{max}+I_{min})$}. When
changing magnetic field $B$ the QPC transmission has to be
readjusted to 1/2, because it sensitively depends on $B$. In some
cases resonances in the QPC transmission characteristics occurred.
In this case the half transmission point with the highest~$\nu_{I}$
was selected. \cite{litvin2} In order to relate the magnetic field
to the filling factor in the MZI arms, the two point conductance of
the interferometer between terminal S and D2 with all QPCs opened
was measured (Fig.2a). The electron density of the narrowest section
of the interferometer can be probed in this way. For the larger
interferometer the transition between the $\nu$=2 and $\nu$=1 Hall
plateaus is located at $B$=5T and corresponds to the filling factor
$\nu$=1.5. For the smaller interferometer the transition point
shifts to 4.8~T. These reference points were used to determine the
filling factor in the range of 3~T$<B<8$~T. The visibility of the
larger MZI shown in Fig.2b with full squares emerges at
$B\gtrsim$4.4T ($\nu$$\approx$1.7), reaches a maximum of 17\% at
$\nu$=1.5, and then non-monotonically decreases to 6\% at $B$=7.5T
($\nu$=1.0). For the smaller interferometer, a measurable visibility
(full circles and empty squares at Fig.2b) emerges near $B$=3.5T
($\nu$$\approx$2.0), reaches maximum around B=5T, and then decreases
to zero at about $B$=8T ($\nu$$\approx$0.9). Right after the
magnetic field is ramped to the next data point, the visibility
drops significantly (about 50\%). However, after some waiting time
it recovers and approaches a saturation value. The saturation takes
up to 10~h. As a reasonable compromise between optimum visibility
and reasonable data acquisition time we have chosen 1 h (full
circles in Fig.2b) and 4 h (open squares in Fig.2b) for the waiting
time. A measurable visibility was observed in the $\nu$ interval
$1.0\lesssim\nu\lesssim2.0$ with a maximum in $\nu_{I}(B)$ near
($\nu$=1.5).
%
\begin{figure}
\includegraphics[width=85mm]{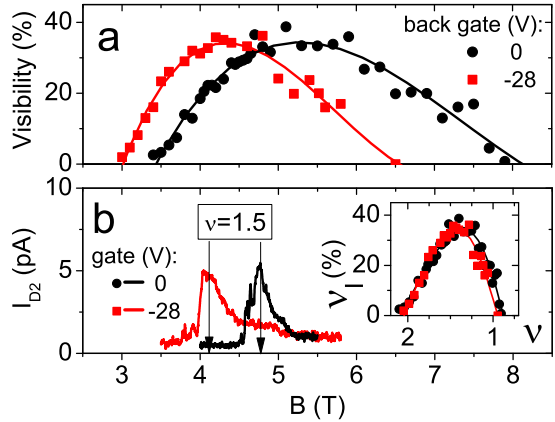}
\caption{(a) Visibility versus magnetic field for two back gate
voltages, i.e., electron densities. (b) Determination of the
$\nu$=1.5 point. Current recorded at detector D2, when QPC1 is
closed and QPC2 is opened. The peak in each curve corresponds to
maximum scattering between edges in the lower MZI arm and correspond
to $\nu$=1.5. Inset: Data from (a) plotted vs filling factor.}
\end{figure}

To assure that it is the filling factor which controls the
visibility and not the absolute magnetic-field magnitude, we changed
the electron density in the interferometer by a back gate. A
decrease in the electron density shifts the entire $\nu_{I}(B)$
curve to smaller magnetic fields (Fig.3a, inset in Fig.3b). As an
independent check of the determination of filling factor within the
interferometer arm, we looked at the backscattering properties. The
quantum point contact QPC0 is set to transmit only the outer
channel, QPC1 is completely closed, and QPC2 is completely opened.
In absence of backscattering within the lower interferometer arm,
QPC1 redirects all current from S to the detector D1, while D2 sees
zero signal. However, when the filling factor is close to 1.5,
backscattering occurs between the counter-propagating lower arm edge
channels which appears as signal in D2 (Fig.3b). The maximum of this
signal corresponds to the point $\nu$=1.5 where the backscattering
is strongest. We see that a back gate voltage of -28~V shifts the
peak, in other words, the point of maximum scattering, from 4.8 to
4.1~T, i.e., the density decreases by 15~\%. The maximum of
visibility was earlier reported to occur on the upper end of the
$\nu$=2 plateau, \cite{roull2} while we observed it at $\nu$=1.5.
This discrepancy may be caused by an unequal filling factor in the
two MZI arms in Ref.~\onlinecite{roull2}. According to
Ref.~\onlinecite{chklovskii} a mesa width which is not much larger
than the depletion length results in different electron densities
and therefore different filling factors. To avoid this problem we
used the same width for both interferometer arms as well as for the
input and output leads.
%
\begin{figure}
\includegraphics[width=85mm]{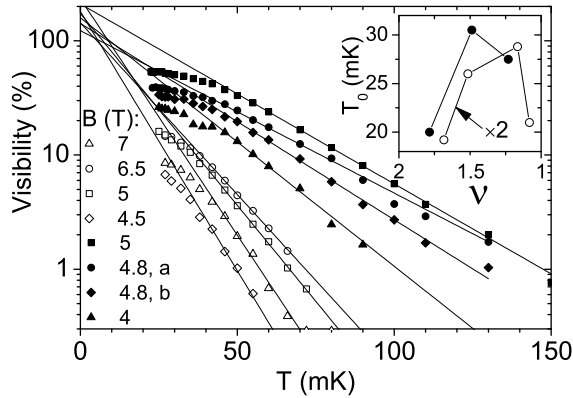}
\caption{(a) Temperature dependencies of visibilities for large
(open symbols) and small (full symbols) MZIs at different magnetic
fields. The curves "a" and "b" for B=4.8T correspond to different
QPC half transmission points. Inset: Characteristic temperatures
$T_{0}$ extracted from the exponential fits according to
Eq.~\ref{1_over_T}. The data for the large interferometer (open
circles) have been multiplied by 2.}
\end{figure}

\section{Characteristic energy scales}
There are three fundamental reasons for a reduction of the
visibility: (i) genuine decoherence by inelastic scattering; (ii)
phase averaging, due to a finite energy window, imposed by
temperature and voltage; \cite{chung} and (iii) phase averaging due
to fluctuations of charges trapped nearby. \cite{marquardt} The last
possibility can probably be excluded, since fluctuations in the
environment are not expected to strongly depend on $B$. A recent
experimental study of $\nu_I(T,B)$ showed that it changes
exponentially with $T$:

\begin{equation}\label{1_over_T}
\nu_{I}=\nu_{I0}\;\exp\left(-T/T_{0}\right)
=\nu_{I0}\;\exp\left(-2L/l_{\varphi}\right)\;,\end{equation} where
the characteristic temperature $T_{0}$ is inversely proportional to
the length of the interferometer arm $L$, i.e., $T_{0}\propto
l_{\varphi}T/2L$ or $l_{\varphi}\propto 2LT_{0}/T$. \cite{roull2}
The electromagnetic environment of the interferometer is expected to
give rise to a $l_\varphi\propto 1/T$ dependence. \cite{seelig} On
the other hand, it is again unclear why such an environmental effect
should strongly vary with $B$.

As demonstrated by the solid lines in Fig.~4, our data also vary
exponentially with $T$ above 45~mK. However, at lower temperatures a
crossover to a weaker temperature dependence is observed. The
presence of such a crossover is reflected by the extrapolated values
of $\nu_{I0}$, which significantly exceed the allowed maximum of
100\%, and the fact that the fit lines do not cross at $T$=0, but
rather at 7~mK. Although it is notoriously hard to exclude that
electron heating contributes to the apparent saturation of $\nu_I$
at $T\lesssim\;$45~mK (Fig.4), the latter two facts refer to the
high-temperature regime and indicate that the behavior of $\nu_I(T)$
may be more complex than a simple exponential.

Despite these differences, our data confirm a clear correlation
between the extracted values of $T_0$ and the visibility, when the
magnetic field is varied.  Like the visibility, $T_0$ has a maximum
around $\nu=1.5$ (see the inset of Fig.~4). \cite{footnote} Thus, it
is clear that the phase coherence length is magnetic field
dependent.

\begin{figure}
\includegraphics[width=85mm]{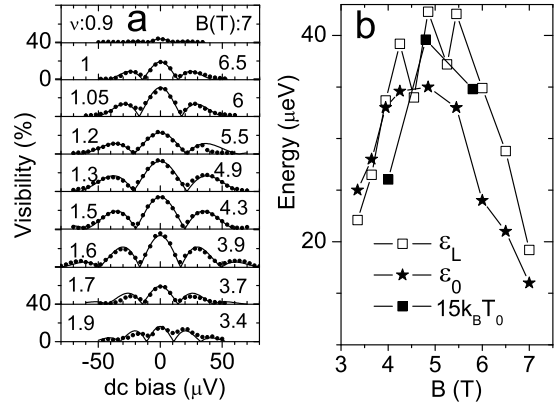}
\caption{Differential visibility of the small MZI. (a) Lobe
structures at different magnetic fields at $V_{BG}=-25\,$V. (b)
Characteristic energies $\varepsilon_L$ and $\varepsilon_{0}$ (see
text) extracted from $\nu_I(V)$ together with $k_BT_0$ as a function
of magnetic field $B$.}
\end{figure}

The visibility can also be measured as a function of a dc voltage
added to the small ac bias. This {\it differential} visibility shows
a lobe structure \cite{neder,roull} vs~$V_{dc}$. In Fig.~5a we show
the evolution of the lobe structure with filling factor. The
observed change of the lobe characteristics
 resembles the change of visibility with $\nu$ (see Fig.~2b),
i.e., the largest distance between the zeros in $\nu_I$ is found
near $\nu$=1.5 and reduces, when $\nu$ moves from =1.5 to $\nu$=1.0
or 2.0. (Fig.5b, open squares). For $\nu>1.5$ more than one pair of
side lobes can be observed (Fig.5a, $B$=3.9T). The data in Fig.~5a
can be well approximated by a product of an oscillatory function and
a Gaussian envelope:
$\nu_{I}=\nu_{I0}\,\left|\cos\left(\pi
eV/\varepsilon_{L}\right)\right|\,\exp\left(-(eV)^{2}/2\varepsilon_{0}^{2}\right),$
which contains the parameters $\varepsilon_L$ as period of the
cosine term and $\varepsilon_0$ as characteristic width of the
envelope. For a direct comparison, we plot these characteristic
energies together with the characteristic temperature $k_BT_0$
extracted from the temperature dependence of $\nu_I$. The energies
$\varepsilon_L$ and $\varepsilon_0$ agree rather well for
$2>\nu>1.5$, while $\varepsilon_0$ is slightly smaller than
$\epsilon_L$ for $\nu<1.5$. On the other hand, $k_BT_0$ is about 15
times smaller. Despite the difference in numbers, the overall
$\nu$~dependence of all energy scales is quite similar. Hence, it
appears that the B~dependence of the characteristic energies is
closely related to the evolution of the structure of edge states
with $\nu$.

\section{Discussion}
For the following, one has to keep in mind that QPC0 is tuned such
that current is injected exclusively into the {\it outer}
incompressible strip with $\nu=1$. According to accepted theory,
\cite{prange} at small bias the current flows close to the interface
between compressible and incompressible stripes. Already in
Ref.~\onlinecite{neder} it was realized that the lobe pattern cannot
be understood in terms of a single electron picture, nor within a
simple mean-field approach. Nevertheless, the charges injected into
the interferometer by the dc~component of the bias appear to induce
an overall phase shift, i.e., they act similar to the modulation
gate. Youn et al. \cite{sim} proposed that the average phase shift
$\langle\delta\rangle$ is determined by the average number $N$ of
non-equilibrium electrons and the intra~channel Coulomb interaction
constant $U_0$:
\begin{equation}
\langle\delta\rangle=N\;\frac{U_0t_{fl}}{\hbar}=\frac{|e|Vt_{fl}}{\hbar}\;\frac{U_0t_{fl}}{\hbar}\;,
\end{equation}
where $t_{fl}=L/v_D$ is the traversal time of the electrons through
the interferometer and $v_D$ the drift velocity along the edge.
Using $v_D\approx3\cdot10^4\,$m/s and $L=9\;\mu$m, one obtains
$N\approx 1$ at $V_{dc}=10 \mu$V. With increasing $N\propto V_{dc}$
the number of possible charge-density distributions in the
interferometer and hence the fluctuations of $\delta$ increases,
which according to the numerical calculations in
Ref.~\onlinecite{sim} leads to a suppression of the visibility at
higher $V_{dc}$. For smaller $U_0$ a larger $V_{dc}$ is needed to
reach the first zero of $\nu_{I}$.

If one now assumes that the interaction parameter $U_0$ is affected
by the screening properties of the environment of the outer edge
channel, the changes of the structure of the 2DEG enclosed by the
outer edge between $\nu=2$ and $\nu=1$ would indeed suggest an
nonmonotonic variation of $U_0(B)$, because the screening is most
effective at $\nu=1.5$, where the entire bulk of the 2DEG is
compressible. On the other hand, at $\nu=1$ and 2, the bulk of the
2DEG is incompressible, implying reduced screening and
correspondingly larger values of $U_0$. This scenario is consistent
with the observed nonmonotonic variation of $\varepsilon_L$. The
energy scales $\varepsilon_L$ and $\varepsilon_0$ must then be
related to $\hbar/t_{fl}$ and $U_0(B)$.

Very recently, another theory based on the chiral Luttinger liquid
approach to the QHE has been suggested. \cite{eugene} In this
theory, the excitations are dipolar (neutral) and charged edge
magnetoplasmon modes with different group velocities $v$ and $u$.
The model allows to calculate $l_\varphi(T)=\hbar uv/\pi k_BT(u-v)$,
which is consistent with the observed $T$~dependence of $l_\varphi$.
In addition it predicts for the ratio between
$\epsilon_L/k_BT_0=2\pi^{2}\simeq 19.7$. From our data we deduce a
similar experimental value of $\epsilon_L/k_BT_0\simeq 15$. This
model requires two well defined edge channels and is valid at
$\nu\geq 1.5$. Hence, it remains to be explained, why the
experimental data show a visibility maximum for $\nu\gtrsim 1.5$.

Another theory for $\nu=1$ results in a $l_\varphi\propto T^{-3}$
for screened and $l_\varphi\propto T^{-1}\ln^2 T$ for unscreened
Coulomb interactions. \cite{gefen} For the transition between
$\nu=1$ and $\nu=2$, so far no theory exists. Qualitatively, one may
expect that a fermionic picture is more appropriate here, since the
screening by the compressible interior of the interior of the 2DEG
tends to suppress the Luttinger liquid effects.

For a better microscopic understanding of the effect, it is
essential to relate the various phenomenological energy scales of
the different theories to a realistic model of the structure of the
edge channels, i.e. the distribution of compressible and
incompressible strips at the mesa edge. \cite{afif2} This
distribution considerably changes between $\nu=1$ and $2$. At
$\nu=1.5$ and above the outer incompressible strip is well localized
at the mesa edge. Depending on the steepness of the confining
potential, it starts to spread out for $\nu\lesssim 1.5$ and fills
the whole mesa, once $\nu=1$ is reached. The precise reason for the
decay of the interference for $\nu<1.5$ is still an open question.
The measured temperature dependence of the visibility suggests that
$l_\varphi$ is suppressed in this regime again. On the other hand,
an edge strip that spreads out over 100 nm and more cannot be
considered as a quasi-one-dimensional object anymore. In particular,
it does not enclose a well defined magnetic flux. The integration of
such QHE-specific features into models considering simple
one-dimensional conduction channels provides new challenges for the
theory.

\section{Conclusions}
In summary we investigated the effect of the filling factor on the
visibility in a Mach-Zehnder interferometer. Surprisingly, the
visibility was found to be highest around $\nu$=1.5 and decreases to
zero when the adjacent integer filling factors are reached. This
dependence originates from an evolution of the structure of edge
channels with magnetic field, which strongly affects three energy
scales of the interference in a very similar way: $k_BT_0$, which
determines the temperature dependence of the visibility in the
linear regime, and $\epsilon_L$ and $\epsilon_0$, which determine
the size and damping of the lobe structure of $\nu_I$ in the
nonlinear regime. This observation suggests that the linear and the
nonlinear regimes are governed by the same energy scale.

\section*{Acknowledgement} We thank M. Heiblum and I. Neder for support
in the framework of EU-Transnational Access Program, Contract
No.~RITA-CT-2003-506095. We are grateful to A. Siddiki, S. Ludwig,
E.V. Devyatov, E.~Sukhorukov and M. B\"{u}ttiker for helpful
discussions. The work was funded by the DFG within the SFB631 "Solid
state quantum information processing" and the BMBF via Project
No.~01BM465 within the program "Nanoquit".


\begin{thebibliography}{15}
\bibitem{footnote} The uppermost trace in Fig.~4 was measured with back gate voltage of
-25~V, corresponding to a reduced electron density.
\end{thebibliography}
\end{document}